\documentclass[12pt]{article}
\usepackage{latexsym,graphicx}
\usepackage{float}
\usepackage{amssymb}
\usepackage{amsmath}
\usepackage{amscd}
\usepackage{color}
\usepackage[left=2cm,top=2.5cm,right=2.5cm,bottom=1.5cm]{geometry}
\usepackage{hyperref}
\usepackage{epstopdf}

\begin{document}
	
	\begin{center}
		\large{\bf{Observational constraints for an axially symmetric transitioning model with bulk viscosity parameterization}} \\
		\vspace{10mm}
		\normalsize{ Archana Dixit$^1$, Anirudh Pradhan$^2$, Vinod Kumar Bhardwaj$^3$, A. Beesham$^4$  }\\
		\vspace{5mm}
		
		\normalsize{$^{1,3}$Department of Mathematics, Institute of Applied Sciences and Humanities, GLA University, Mathura-281 406, Uttar Pradesh, India}\\
			\vspace{2mm}
			\normalsize{$^{2}$ Centre for Cosmology, Astrophysics and Space Science (CCASS), GLA University,\\
			Mathura-281 406, Uttar Pradesh, India}\\
		\vspace{2mm}
		\normalsize{$^{4}$Department of Mathematical Sciences, University of Zululand Private Bag X1001 Kwa-Dlangezwa 3886 South Africa}\\
		\vspace{2mm}
		$^1$E-mail:archana.dixit@gla.ac.in\\
		\vspace{2mm}
		$^2$E-mail:pradhan.anirudh@gmail.com \\
		\vspace{2mm}
		$^3$E-mail:dr.vinodbhardwaj@gmail.com\\
			\vspace{2mm}
		$^{4}$E-mail: abeesham@yahoo.com \\
		
			\vspace{10mm}
		
	\end{center}

\begin{abstract}
   In this paper, we have analyzed the significance of bulk viscosity in an axially symmetric Bianchi type-I model to study the accelerated expansion of the universe. We have considered four bulk viscosity parameterizations for the matter-dominated cosmological model. The function of the two significant Hubble $H(z)$ and deceleration parameters are discussed in detail. The energy parameters of the universe are computed using the most recent observational Hubble data (57 data points) in the redshift range $0.07 \leq z \leq 2.36)$. In this model, we obtained all feasible solutions with the viscous component and analyzed the universe's expansion history. Finally, we analyzed the statefinder diagnostic and found some interesting results. The outcomes of our developed model now properly align with observational results.\\
	
\end{abstract}


\section{Introduction}
The recent finding of the universe's accelerated expansion has been discussed as the most well-known issue in modern cosmology. Different observational data, such as Type Ia Supernovae (SNIa)\cite{ref1,ref2}, Cosmic Microwave Background Radiation (CMB) \cite{ref3,ref4,ref5} and gravitational lensing \cite{ref6,ref7}, have all contributed to confirming the accelerating expansion. This phenomenon is explained in the context of general relativity (GR) by introducing an unidentified energy source known as dark energy (DE). Dark energy produces a gravitational repulsion that aids in accelerating motion by exerting a strong negative force that results in an anti-gravity effect.
In the FRW model, viscosity is the single observable phenomenon in the study of cosmological models with viscous fluid. Common knowledge holds that bulk viscosity causes the inflationary phase and generates a negative pressure resembling repulsive gravity. \\

Fluid mechanics has a concept called viscosity. It is divided into two types: shear viscosity and bulk viscosity, and it is related to the velocity gradient. Shear viscosity is important in viscous cosmology because it relates to spacetime anisotropy. The energy-momentum tensor (EMT) can have a bulk viscosity term in the Friedmann-Robertson-Walker (FRW) framework of at most ($\rho-\xi\theta$), where $\xi$ is the bulk viscosity and $theta$ is the expansion scalar. Also, bulk viscosity and the grand-unified-theory phase transition may help explain why the universe is expanding faster than expected. The bulk viscosity in cosmology has been explored in \cite{ref14, ref15,ref16,ref17,ref18,ref19} in many aspects. Numerous cosmologists have examined the influence of bulk viscosity in the cosmic media on the DE phenomena  \cite{ref20,ref21}. In this context, Santos et al. obtained perfect solutions to the isotropic homogeneous model, where the bulk viscosity was determined to be a power function of the energy density. An inflationary cosmology may develop due to the bulk viscosity related to the grand unified theory (GUT). Cataldo et al. \cite{ref22} showed the phantom models with viscous fluid. Carlevaro and Montani \cite{ref23} examined the consequences of bulk viscosity being produced on the stability of the very early universe. In the same direction,  authors \cite{ref24} work on viscous cosmology in the Weyl-type f(Q, T) gravity in the same framework. Current findings from Goswami  \cite{ref25}, and Kotambkar  \cite{ref26} have focused on simulating the accelerating and dynamical behavior of Chaplygin gas, as well as cosmological and gravitational ``constants" in various scenarios.\\

In cosmology, kinematic variables like the Hubble parameter (H) and the deceleration parameter (q) are all derived from the scale factor's derivatives and may also be used to examine cosmic acceleration (for details, see Sect. 2). These redshift transitions were discussed in \cite{ref27,ref28}. The parameters $\Omega_m$, $\Omega_{de}$ and present value of Hubble constant $H_{0}$ have been estimated using latest OHD data set \cite{ref29,ref30,ref31,ref33,ref34,ref35,ref36,ref37,ref38,ref39,ref39a,ref39b,ref40,ref41} and results are found to be matched with the results obtained from SNIa data, BAO, CMB observations and also with statistical analysis compilation by Huchra's $H_{0}$ \cite{ref42}.\\

In this context, the Hubble parameter data set and the cosmic expansion rate are the function of redshift $z$, and $ H(z)$ is especially fascinating in this direction. Current H(z) data contains a higher redshift range  than Type Ia supernovae,  $0.07 \leq z \leq 2.36$ \cite{ref43,ref44,ref45,ref46,ref47}.
The H(z) data can also be used to identify the cosmological deceleration-acceleration transition  \cite{ref48,ref49}; measure the Hubble constant $H_{0}$ \cite{ref50,ref51,ref52}; and constrain the standard cosmological parameters, such as the non-relativistic matter and dark energy density parameters. Additionally, spatial curvature may be constrained using H(z) measurements in combination with distance-redshift data (Clarkson et al. 2007, 2008) \cite{ref53,ref54}.\\
					
 The most effective and straightforward models that fully describe the anisotropic effects are those of the Bianchi type. These anisotropic models have several advantages over isotropic FRW models, including the ability to describe the history of the early universe and help to develop more generic cosmological models \cite{ref54a,ref55,ref56,ref57}. Locally rotational symmetric (LRS) space-time is a subclass of Bianchi type I models, which is spatially homogeneous and represent symmetry along a certain direction. A significant number of contemporary accelerating universe models on LRS spacetime have been published in the literature \cite{ref58,ref59,ref60,ref61,ref62,ref63}. 
In the context of viscosity, we have shown how an axially symmetric Bianchi type-I model of the cosmos evolved in this study. Our model predicts that the universe has just undergone a deceleration and is now accelerating. Roles of the two crucial deceleration $q(z)$ and Hubble $H(z)$ parameters are examined. The most recent observational Hubble data determine the universe's energy parameters (57 data points). 
Through statefinder diagnostics, we have also discussed the model's stability analysis. 
The investigation shows that the model is a quintessence type in the late time and leads toward the $\Lambda$CDM model. The present model is adequately consistent with empirical results.\\

The manuscript is organized as follows: In section 1, we discussed the present cosmological scenario. In section 2, The model's field equations are present in section 2. In section 3, we have discussed the evolutionary trajectories. Results and discussions are mentioned in section 4. The last section, 5, is devoted to conclusions.

 
\section{  Field Equations of the model }
We take into account the form's LRS Bianchi type I metric.
 	
\begin{equation}
 	\label{1}
 ds^2=A (dx)^2+ B^2 \left( dy^2+ dz^2 \right)-dt^2
 	\end{equation}
 	Here, ``A'' and ``B" are the time-dependent metric functions. 
 	The Einstein field equation for general relativity with the cosmological constant ($\Lambda$) is:
 	\begin{equation}
 		\label{2}
 	\Lambda g_{ij}+	R_{ij}-\frac{1}{2} g_{ij} R =-T_{ij}
 \end{equation}
Considering the EMT in the form  $ T_{ij} = (\bar p+\rho_{m}) u_{i} u_{j}-\bar p  g_{ij} $, where $\rho_{m}$ is the matter energy density and the bulk pressure $ \bar p $ is given as

\begin{equation}
	\label{3}
	\bar p= p- 3 \xi H
\end{equation}
where $p$ is normal pressure that is equal to 0 for the nonrelativistic matter. The aforementioned EMT is comparable to perfect but has an equivalent pressure of $\bar p$.
This pressure is created by adding the viscous pressure ($p_{vis} =-3\xi H$) to the fluid pressure ($p$). As mentioned above, the viscous pressure  $p_{vis}$ can be considered a measurement of the pressure to restore the local thermodynamics equilibrium. We conclude that in the presence of a viscous fluid, $p=0$ governs the universe of pressureless matter.\\

The field equations for the metric  (\ref{1}) are expressed as:
\begin{equation}
	\label{4}
2 \frac{\ddot{B}}{B}+ \frac{\dot{B}^2}{B^2}=-\bar p+\Lambda
\end{equation}
\begin{equation}
	\label{5}
\frac{\ddot{A}}{A}+\frac{\ddot{B}}{B}+ \frac{\dot{A} \dot{B} }{A B}=-\bar p+\Lambda
\end{equation}
\begin{equation}
	\label{6}
2\frac{\dot{A} \dot{B}}{AB}+\frac{\dot{B}^2}{B^2}=\rho_{m}+\Lambda
\end{equation}
To proposed the explicit solution of the field equations, we have assumed the scale factor as $ a =\left(A B^{2}\right)^{1/3} $, volume as  $ V =A B^{2} $ and Hubble's parameter as $ H = \frac{1}{3} \left(\frac{\dot{A}}{A}+2\frac{\dot{B}}{B}\right)= \frac{\dot{a}}{a} $.\\
From    Eqs. (\ref{4})-(\ref{5}), we have
\begin{equation}
\label{7}
\frac{d}{dt}\left(\frac{\dot{A}}{A}-\frac{\dot{B}}{B}\right)=-\left(\frac{\dot{A}}{A}-\frac{\dot{B}}{B}\right) \left(\frac{\dot{A}}{A}+2\frac{\dot{B}}{B}\right)
\end{equation}
On integration, Eq. (\ref{7}) we get
\begin{equation}
	\label{8}
\frac{\dot{A}}{A}-\frac{\dot{B}}{B} =\frac{c}{a^3}
\end{equation}
Also, we have
\begin{equation}
	\label{9}
\frac{\dot{A}}{A}+2\frac{\dot{B}}{B} =3\frac{\dot{a}}{a}
\end{equation}
Solving Eqs. (\ref{8}) and (\ref{9}), we get
\begin{equation}
	\label{10}
\frac{\dot{A}}{A}=\frac{\dot{a}}{a}+\frac{2}{3} \frac{c}{a^3}
\end{equation}
\begin{equation}
	\label{11}
\frac{\dot{B}}{B}=\frac{\dot{a}}{a}-\frac{1}{3} \frac{c}{a^3}
\end{equation}
Using Eqs. (\ref{10}) and (\ref{11}) in Eq.(\ref{6}), we get
\begin{equation}
	\label{12}
H^2 =\frac{1}{3}\left(\rho_{m}+\Lambda+\frac{1}{3} \frac{c^{2}}{a^6}\right)
\end{equation}
 The energy conservation law, $ \dot{\rho_{m}}+3 H (p_{m}+\rho_{m}) =0$, holds for barotropic matter. For the dust filled current universe ``$ p_{m}=0 $ and $ \rho_{m} \propto a^{-3} $". Using the relation $ \frac{a_{0}}{a} =1+z $, we get $ \rho_{m} = \rho_{m0} (1+z)^3 $. The term $\frac{1}{3} \frac{c^{2}}{a^6} =\rho_{\sigma0} (1+z)^{6} =\rho_{\sigma}$ denotes the anisotropy energy density. For the dust filled universe ($ p_{m}=0 $), the density parameters are defined as $ \Omega_{m} = \frac{\rho_{m}}{\rho_{c}}  $ and $ \Omega_{\sigma} = \frac{\rho_{\sigma}}{\rho_{c}} $, where $ \rho_{c} = \frac{3 H^{2}}{8 \pi G} $ \& $ 8 \pi G\approx 1 $. Thus, Eq. (\ref{12}) can also be witten as
\begin{equation}
	\label{13}
H^2 = H^{2}_{0}\left[ (1+z)^{3}\Omega_{m0}+\Omega_{\Lambda0}+ (1+z)^{6}\Omega_{\sigma0}\right]
\end{equation}

For $ z=0 $, the relationship between energy parameters is obtained from Eq. (\ref{13}) as:
\begin{equation}
	\label{14}
1=\Omega_{m0} +\Omega_{\Lambda0}+\Omega_{\sigma0} 
\end{equation}

The DP for the model can be expressed as:
\begin{equation}
	\label{15}
	q= -\frac{a \ddot a }{\dot a^{2}}=--\frac{ \ddot a }{ a}\frac{1 }{ H^{2}}
\end{equation}

\begin{equation}
	\label{16}
	\frac{\ddot a }{ a}=\frac{1}{2}\left[3 \xi H +3H_{0}^{2}\Omega_{\Lambda_{0}}-H^{2}-2H_{0}^{2}\Omega_{\sigma_{0}}(1+z)^{6}+3Hc(1+z)^{3}-2c(1+z)^{4}\right]
\end{equation}
By using Eqs. (\ref{15}) and (\ref{16}), we obtained the deceleration parameter as:
\begin{equation}
		\label{17}
	q= \frac{1}{2}\left[\frac{(3\xi+2c(1+z)^{3})}{H} +\frac{3H_{0}^{2}(\Omega_{\Lambda_{0}}-\Omega_{\sigma_{0}}(1+z)^{6})}{H^{2}}-\frac{2c(1+z)^{4}}{H^{2}}-1\right]
\end{equation}
By using Eq. (\ref{15}) the deceleration parameter can also be rewritten as:
\begin{equation}
		\label{18}
	q= \frac{1}{2}\left[\frac{3(\xi+c(1+z)^{3})}{\sqrt{[(1+z)^{3}\Omega_{m0}+\Omega_{\Lambda0}+ (1+z)^{6}\Omega_{\sigma0}}} +\frac{3(\Omega_{\Lambda_{0}}-\Omega_{\sigma_{0}}(1+z)^{6}-2c(1+z)^{4})}{\left[ (1+z)^{3}\Omega_{m0}+\Omega_{\Lambda0}+ (1+z)^{6}\Omega_{\sigma0}\right]}-1\right]
\end{equation}

$H$ plays a crucial role in the explanation of the universe's expansion and is also significantly helpful in estimating the age of the cosmos. In the similar way, the deceleration parameter explains the change in phase (acceleration or deceleration) that occurs during the evolution of the cosmos.\\  


\section{State-finders}
In this section, we have focused on analyzing the statefinder diagnosis.
Statefinder Diagnostic (SFD) is a geometrical parameter pair approach developed for the broad analysis of many dark energy models \cite{ref64,ref65}. The pair is designated by ${r, s}$ where $r$ and $s$ are defined as:
 
\begin{equation}
	\label{19}
	r= 2q^{2}+q+\frac{\dot{q}}{H}
\end{equation}

\begin{equation}
	\label{20}
	s= \frac{r-1}{3 (q-\frac{1}{2})}
\end{equation}

The various evolutionary trajectories of the geometric pair $({r~s})$ developing in the $(r-s)$ plane allow for the examination of several dark energy scenarios. Here, the two well-known geometrical variables that characterize the history of the universe's expansion are the Hubble parameter $H$, which represents the rate of the universe's expansion, and the deceleration parameter $q$, which represents the rate of acceleration/deceleration of the expanding cosmos.
The standard $\Lambda$ CDM model of cosmology is represented by the fixed point ${(r, s)} = {(1, 0)}$ while the standard matter-dominated Universe, SCDM, is represented by the fixed point ${(r, s)} = {(1, 1)}$. This is a symbolic part of the SFD.  Besides the $\Lambda$ CDM and SCDM models, the SFD analysis can successfully explain the different dark energy candidates. Quintessence, braneworld dark energy models, Chaplygin gas, and a few additional interacting dark energy models are among them. This is completed by finding some particular region in the figure in the distinctive trajectories \cite{ref66,ref67,ref68}.
To explain the behavior of our model, we have used the SFD technique. We have explained the convergence and divergence characteristics and compared the results with $\Lambda$ CDM and SCDM. The four types of bulk viscosity parameterization are used in our derived model, and their physical significance is discussed.


\subsection{Solution with Bulk viscosity }
On thermodynamic grounds, $\xi$ in  Eq. (\ref{16}) is usually a positive number, and it may depend on the cosmic time $t$, the scale factor $a$, or the energy density. Since viscosity may take many forms,  Eq. (\ref{16}) can be solved numerically or precisely. In our derived model following are some options for $\xi$ are examined in particular cases:\\
{\bf Case I:} In this case, we have choice
$\xi=\xi_{0}$ to determine the certain cosmological parameters by using the constant bulk viscosity. We  have calculated the deceleration parameter $q$  and statefinders by using  Eq. (\ref{18}) and the relation $\xi=\xi_{0}$,
\cite{ref15,ref69} we obtained decelerating parameter as :


\begin{equation}
		\label{21}
	q= \frac{1}{2}\left[\frac{(3\xi_{0}+2c(1+z)^{3})}{\sqrt{[(1+z)^{3}\Omega_{m0}+\Omega_{\Lambda0}+ (1+z)^{6}\Omega_{\sigma0}}} +\frac{3(\Omega_{\Lambda_{0}}-\Omega_{\sigma_{0}}(1+z)^{6}-2c(1+z)^{4})}{\left[ (1+z)^{3}\Omega_{m0}+\Omega_{\Lambda0}+ (1+z)^{6}\Omega_{\sigma0}\right]}-1\right]
\end{equation}
The statefinder $(r-s)$ can be obtained, by using  Eqs. (\ref{19}), (\ref{20}) and  (\ref{21})
\begin{eqnarray}
		\label{22}
	r&=&\bigg[\frac{H_0^2}{H^4(z)}\bigg((-z-1) \left(3 (z+1)^2 \Omega _{\text{m0}}+6 (z+1)^5 \Omega _{\text{$\sigma $0}}\right)\bigg)\nonumber\\
	&\times& \frac{1}{2}\bigg(2 c H (z) (z+1)^3-2 c (z+1)^4-H^2 (z)+3 H_0^2 (\Omega _{\text{$\Lambda $0}}- (z+1)^6 \Omega _{\text{$\sigma $0}})+3 H(z) \xi_{0} \bigg)\nonumber\\
	&+& \frac{1}{H^4(z)} \bigg(2 c H (z) (z+1)^3-2 c (z+1)^4-H^2 (z)+3 H_0^2 (\Omega _{\text{$\Lambda $0}} - (z+1)^6 \Omega _{\text{$\sigma $0}})+3 H (z)\xi_{0} \bigg){}^2\nonumber\\
	&-&\frac{1}{2 H^2(z)}\bigg(2 c H (z) (z+1)^3-2 c (z+1)^4-H^2 (z)\nonumber\\
	&+&3 H_0^2 (\Omega _{\text{$\Lambda $0}}- (z+1)^6 \Omega _{\text{$\sigma $0}})+3 H (z) \xi_{0}\bigg) \bigg]
\end{eqnarray}

\begin{eqnarray}
		\label{23}
	s&=&\bigg[\frac{2 H(z)^{2}}{3 \big(-2 c H z (z+1)^3+2 c (z+1)^4-3 H_0^2 \Omega _{\text{$\Lambda $0}}-3 H \xi  z+3 H_0^2 (z+1)^6 \Omega _{\text{$\sigma $0}}\big)}\bigg]\nonumber\\
&\times&\bigg[-1+\biggl\{\frac{H_0^2}{H^4(z)}\bigg((-z-1) \left(3 (z+1)^2 \Omega _{\text{m0}}+6 (z+1)^5 \Omega _{\text{$\sigma $0}}\right)\bigg)\nonumber\\
&\times& \frac{1}{2}\bigg(2 c H (z) (z+1)^3-2 c (z+1)^4-H^2 (z)+3 H_0^2 (\Omega _{\text{$\Lambda $0}}- (z+1)^6 \Omega _{\text{$\sigma $0}})+3 H(z) \xi_{0} \bigg)\nonumber\\
&+& \frac{1}{H^4(z)} \bigg(2 c H (z) (z+1)^3-2 c (z+1)^4-H^2 (z)+3 H_0^2 (\Omega _{\text{$\Lambda $0}} - (z+1)^6 \Omega _{\text{$\sigma $0}})+3 H (z)\xi_{0} \bigg){}^2\nonumber\\
&-&\frac{1}{2 H^2(z)}\bigg(2 c H (z) (z+1)^3-2 c (z+1)^4-H^2 (z)\nonumber\\
&+&3 H_0^2 (\Omega _{\text{$\Lambda $0}}- (z+1)^6 \Omega _{\text{$\sigma $0}})+3 H (z) \xi_{0}\bigg)\biggr\}\bigg]
\end{eqnarray}


{\bf Case-II}:-In the homogeneous models, $\xi$ only depends on time; therefore, we may consider it a function of the Universe's energy density $\rho$. Citing references \cite{ref16,ref70}, it is stated ``we assume that the bulk viscosity coefficient depends on $\rho$ as a power-law of the form when $\xi=\xi_{0}\rho^{n}$. Where $\xi_{0}>0$ dimensional constant while $n\geq 0$ is a numerical constant". In the case of a radiative fluid with a low density, $n$ may be equal to 1. To achieve realistic findings, it is reasonable to assume that $0\leq n\leq 1/2$, as suggested by Belinskii and Khalatnikov \cite{ref71}. By substituting the relation
$\xi=\xi_{0}$ $\rho^{n}$ in  Eq. (\ref{18}), we have calculate the deceleration parameter as:

\begin{equation}
		\label{24}
	q= \frac{1}{2}\left[\frac{3(\xi_{0}\left(3 H_0^2 (z+1)^3 \Omega _{\text{m0}}\right){}^n+2c(1+z)^{3})}{\sqrt{[(1+z)^{3}\Omega_{m0}+\Omega_{\Lambda0}+ (1+z)^{6}\Omega_{\sigma0}}} +\frac{3(\Omega_{\Lambda_{0}}-\Omega_{\sigma_{0}}(1+z)^{6}-2c(1+z)^{4})}{\left[ (1+z)^{3}\Omega_{m0}+\Omega_{\Lambda0}+ (1+z)^{6}\Omega_{\sigma0}\right]}-1\right]
\end{equation}
The statefinder $(r-s)$ can be obtained, by using Eqs.(\ref{19}), (\ref{20}) and  (\ref{24})

\begin{eqnarray}
		\label{25}
	r&=&\bigg[\frac{H_0^2}{H^4(z)}\bigg((-z-1) \left(3 (z+1)^2 \Omega _{\text{m0}}+6 (z+1)^5 \Omega _{\text{$\sigma $0}}\right)\bigg)\nonumber\\
	&\times& \frac{1}{2}\bigg(2 c H (z) (z+1)^3-2 c (z+1)^4-H^2 (z)+3 H_0^2 (\Omega _{\text{$\Lambda $0}}- (z+1)^6 \Omega _{\text{$\sigma $0}})+3^{n+1} H(z) \xi_{0} \left( H_0^2 (z+1)^3 \Omega _{\text{m0}}\right){}^n\bigg)\nonumber\\
	&+& \frac{1}{H^4(z)} \bigg(2 c H (z) (z+1)^3-2 c (z+1)^4-H^2 (z)+3 H_0^2 (\Omega _{\text{$\Lambda $0}} - (z+1)^6 \Omega _{\text{$\sigma $0}})\nonumber\\
	&+&3^{n+1} H (z)\xi_{0}\left(H_0^2 (z+1)^3 \Omega _{\text{m0}}\right){}^n \bigg){}^2
	-\frac{1}{2 H^2(z)}\bigg(2 c H (z) (z+1)^3-2 c (z+1)^4-H^2 (z)\nonumber\\
	&+&3 H_0^2 (\Omega _{\text{$\Lambda $0}}- (z+1)^6 \Omega _{\text{$\sigma $0}})+3^{n+1} H (z) \xi_{0}\left( H_0^2 (z+1)^3 \Omega _{\text{m0}}\right){}^n\bigg) \bigg]
\end{eqnarray}

	\begin{eqnarray}
			\label{26}
		s&=&\bigg[\frac{2 H(z)^{2}}{3 \big(-2 c H z (z+1)^3+2 c (z+1)^4-3 H_0^2 \Omega _{\text{$\Lambda $0}}- 3^{n+1} H(z)\xi_{0} \left( H_0^2 (z+1)^3 \Omega _{\text{m0}}\right){}^n +3 H_0^2 (z+1)^6 \Omega _{\text{$\sigma $0}}\big)}\bigg]\nonumber\\
		&\times&\bigg[-1+\biggl\{\frac{H_0^2}{H^4(z)}\bigg((-z-1) \left(3 (z+1)^2 \Omega _{\text{m0}}+6 (z+1)^5 \Omega _{\text{$\sigma $0}}\right)\bigg)\nonumber\\
		&\times& \frac{1}{2}\bigg(2 c H (z) (z+1)^3-2 c (z+1)^4-H^2 (z)+3 H_0^2 (\Omega _{\text{$\Lambda $0}}- (z+1)^6 \Omega _{\text{$\sigma $0}})+3^{n+1} H(z)\xi_{0} \left( H_0^2 (z+1)^3 \Omega _{\text{m0}}\right){}^n \bigg)\nonumber\\
		&+& \frac{1}{H^4(z)} \bigg(2 c H (z) (z+1)^3-2 c (z+1)^4-H^2 (z)+3 H_0^2 (\Omega _{\text{$\Lambda $0}} - (z+1)^6 \Omega _{\text{$\sigma $0}})+3^{n+1} H(z)\xi_{0} \left( H_0^2 (z+1)^3 \Omega _{\text{m0}}\right){}^n \bigg){}^2\nonumber\\
		&-&\frac{1}{2 H^2(z)}\bigg(2 c H (z) (z+1)^3-2 c (z+1)^4-H^2 (z)\nonumber\\
		&+&3 H_0^2 (\Omega _{\text{$\Lambda $0}}- (z+1)^6 \Omega _{\text{$\sigma $0}})+3^{n+1} H(z)\xi_{0} \left( H_0^2 (z+1)^3 \Omega _{\text{m0}}\right){}^n\bigg)\biggr\}\bigg]
	\end{eqnarray}


{\bf Case III:} In this instance, the bulk viscosity is written as $ \xi = \xi_{0}+\xi_{1}H$, where $\xi_{0}$ and $\xi_{1}$ are traditionally two constants, and the overhead dot denotes a derivative about time.
This bulk viscosity is being considered since, according to fluid mechanics, the transport/viscosity phenomenon involves velocity, which is a function of time. Related to the scalar expansion $\theta = 3\frac{\dot a}{a}$,$\xi =\xi_{0}(constant)$ and $\xi\propto \theta$ are both discussed in prior references  \cite{ref15,ref69}. By using Eq(18) and the relation $ \xi = \xi_{0}+\xi_{1}H$ we obtained the deceleration parameter obtained as:

\begin{equation}
		\label{27}
	q= \frac{1}{2}\left[\frac{(3\xi_{0}+2c(1+z)^{3})}{\sqrt{[(1+z)^{3}\Omega_{m0}+\Omega_{\Lambda0}+ (1+z)^{6}\Omega_{\sigma0}}} +\frac{3(\Omega_{\Lambda_{0}}-\Omega_{\sigma_{0}}(1+z)^{6}-2c(1+z)^{4})}{\left[ (1+z)^{3}\Omega_{m0}+\Omega_{\Lambda0}+ (1+z)^{6}\Omega_{\sigma0}\right]}+3\xi_{1}-1\right]
\end{equation}
The statefinder $(r-s)$ can be obtained, by using  Eqs. (\ref{19}), (\ref{20}) and (\ref{27})

	\begin{eqnarray}
			\label{28}
		r&=&\bigg[\frac{H_0^2}{H^4(z)}\bigg((-z-1) \left(3 (z+1)^2 \Omega _{\text{m0}}+6 (z+1)^5 \Omega _{\text{$\sigma $0}}\right)\bigg)\nonumber\\
		&\times& \frac{1}{2}\bigg(2 c H (z) (z+1)^3-2 c (z+1)^4-H^2 (z)+3 H_0^2 (\Omega _{\text{$\Lambda $0}}- (z+1)^6 \Omega _{\text{$\sigma $0}})+3 H(z) (\xi_{0}+\xi_{1} H(z)) \bigg)\nonumber\\
		&+& \frac{1}{H^4(z)} \bigg(2 c H (z) (z+1)^3-2 c (z+1)^4-H^2 (z)+3 H_0^2 (\Omega _{\text{$\Lambda $0}} - (z+1)^6 \Omega _{\text{$\sigma $0}})+3 H (z) (\xi_{0}+\xi_{1} H(z)) \bigg){}^2\nonumber\\
		&-&\frac{1}{2 H^2(z)}\bigg(2 c H (z) (z+1)^3-2 c (z+1)^4-H^2 (z)\nonumber\\
		&+&3 H_0^2 (\Omega _{\text{$\Lambda $0}}- (z+1)^6 \Omega _{\text{$\sigma $0}})+3 H (z)  (\xi_{0}+\xi_{1} H(z))\bigg) \bigg]
	\end{eqnarray}

	\begin{eqnarray}
			\label{29}
		s&=&\bigg[\frac{2 H(z)^{2}}{3 \big(-2 c H z (z+1)^3+2 c (z+1)^4-3 H_0^2 \Omega _{\text{$\Lambda $0}}-3 H(z) (\xi_{0}+\xi_{1} H(z))+3 H_0^2 (z+1)^6 \Omega _{\text{$\sigma $0}}\big)}\bigg]\nonumber\\
		&\times&\bigg[-1+\biggl\{\frac{H_0^2}{H^4(z)}\bigg((-z-1) \left(3 (z+1)^2 \Omega _{\text{m0}}+6 (z+1)^5 \Omega _{\text{$\sigma $0}}\right)\bigg)\nonumber\\
		&\times& \frac{1}{2}\bigg(2 c H (z) (z+1)^3-2 c (z+1)^4-H^2 (z)+3 H_0^2 (\Omega _{\text{$\Lambda $0}}- (z+1)^6 \Omega _{\text{$\sigma $0}})+3 H(z) (\xi_{0}+\xi_{1} H(z)) \bigg)\nonumber\\
		&+& \frac{1}{H^4(z)} \bigg(2 c H (z) (z+1)^3-2 c (z+1)^4-H^2 (z)+3 H_0^2 (\Omega _{\text{$\Lambda $0}} - (z+1)^6 \Omega _{\text{$\sigma $0}})+3 H(z) (\xi_{0}+\xi_{1} H(z)) \bigg){}^2\nonumber\\
		&-&\frac{1}{2 H^2(z)}\bigg(2 c H (z) (z+1)^3-2 c (z+1)^4-H^2 (z)\nonumber\\
		&+&3 H_0^2 (\Omega _{\text{$\Lambda $0}}- (z+1)^6 \Omega _{\text{$\sigma $0}})+3 H(z) (\xi_{0}+\xi_{1} H(z))\bigg)\biggr\}\bigg]
	\end{eqnarray}


{\bf Case IV:} 
In fourth case, for the parametrization of the bulk viscosity we assume the viscous component  as,$ \xi = \xi_{0}+\xi_{1}H+\xi_{2}(\dot H+H^{2})$. Where $\xi_{0}$, $\xi_{1}$ and $\xi_{2}$ are constants  \cite{ref16} \cite{ref19}. The bulk viscosity coefficient in an expanding universe may be affected by both velocity and acceleration. 
The simplest form is a linear combination of three terms: the first is a constant $\xi_{0}$, the second of which is proportional to the Hubble parameter, which determines how to bulk viscosity varies with velocity, and the third of which is proportional to $\frac{\ddot a}{a}$, which determines the rate of change of bulk viscosity. Using  Eq. (\ref{18}) and the relationship $\xi = \xi_{0}+\xi_{1}H+\xi_{2}(\dot H+H^{2})$, we obtained the  deceleration parameter as:

\[	q= \frac{1}{2}\left[\frac{3(\xi_{0}+\xi_{2}\left(H_0^2 \left(\Omega _{\text{$\Lambda $0}}+(z+1)^3 \Omega _{\text{m0}}+(z+1)^6 \Omega _{\text{$\sigma $0}}\right)-\frac{1}{2} H_0^2 (z+1) \left(3 (z+1)^2 \Omega _{\text{m0}}+6 (z+1)^5 \Omega _{\text{$\sigma $0}}\right)\right)}{\sqrt{[(1+z)^{3}\Omega_{m0}+\Omega_{\Lambda0}+ (1+z)^{6}\Omega_{\sigma0}}}\right]
\]
\begin{equation}
		\label{30}
	+\frac{1}{2}\left[	\frac{2c(1+z)^{3})}{\sqrt{[(1+z)^{3}\Omega_{m0}+\Omega_{\Lambda0}+ (1+z)^{6}\Omega_{\sigma0}}} +\frac{3(\Omega_{\Lambda_{0}}-\Omega_{\sigma_{0}}(1+z)^{6}-2c(1+z)^{4})}{\left[ (1+z)^{3}\Omega_{m0}+\Omega_{\Lambda0}+ (1+z)^{6}\Omega_{\sigma0}\right]}+3\xi_{1}-1\right]
\end{equation}

The statefinder $(r-s)$ can be obtained, by using  Eqs. (\ref{19}), (\ref{20}) and  (\ref{30})

	\begin{eqnarray}
			\label{31}
	r&=&\bigg[\frac{H_0^2}{H^4(z)}\bigg((-z-1) \left(3 (z+1)^2 \Omega _{\text{m0}}+6 (z+1)^5 \Omega _{\text{$\sigma $0}}\right)\bigg)\nonumber\\
	&\times& \frac{1}{2}\bigg(2 c H (z) (z+1)^3-2 c (z+1)^4-H^2 (z)+3 H_0^2 (\Omega _{\text{$\Lambda $0}}- (z+1)^6 \Omega _{\text{$\sigma $0}})\nonumber\\
	&+&3 H(z) \big(\xi_{0}+\xi_{1} H(z)+\xi_{2} (\dot H(z)+ H(z)^{2})\big) \bigg)\nonumber\\
	&+& \frac{1}{H^4(z)} \bigg(2 c H (z) (z+1)^3-2 c (z+1)^4-H^2 (z)+3 H_0^2 (\Omega _{\text{$\Lambda $0}} - (z+1)^6 \Omega _{\text{$\sigma $0}})\nonumber \\
	&+&3 H(z) \big(\xi_{0}+\xi_{1} H(z)+\xi_{2} (\dot H(z)+ H(z)^{2})\big) \bigg){}^2\nonumber\\
	&-&\frac{1}{2 H^2(z)}\bigg(2 c H (z) (z+1)^3-2 c (z+1)^4-H^2 (z)\nonumber\\
	&+&3 H_0^2 (\Omega _{\text{$\Lambda $0}}- (z+1)^6 \Omega _{\text{$\sigma $0}})3 H(z) \big(\xi_{0}+\xi_{1} H(z)+\xi_{2} (\dot H(z)+ H(z)^{2})\big)\bigg) \bigg]
\end{eqnarray}

\begin{eqnarray}
		\label{32}
	s&=&\bigg[\frac{2 H(z)^{2}}{3 \big(-H(z)(2c(z+1)^3+3 H(z) (\xi_{0}+\xi_{1} H(z)+\xi_{2} (\dot H(z)+ H(z)^{2}))+2 c (z+1)^4-3 H_0^2 (\Omega _{\text{$\Lambda $0}}+ (z+1)^6 \Omega _{\text{$\sigma $0}})\big)}\bigg]\nonumber\\
	&\times&\bigg[-1+\biggl\{\frac{H_0^2}{H^4(z)}\bigg((-z-1) \left(3 (z+1)^2 \Omega _{\text{m0}}+6 (z+1)^5 \Omega _{\text{$\sigma $0}}\right)\bigg)\nonumber\\
	&\times& \frac{1}{2}\bigg(2 c H (z) (z+1)^3-2 c (z+1)^4-H^2 (z)\nonumber\\
	&+&3 H_0^2 (\Omega _{\text{$\Lambda $0}}- (z+1)^6 \Omega _{\text{$\sigma $0}})+3 H(z) (\xi_{0}+\xi_{1} H(z)+\xi_{2} (\dot H(z)+ H(z)^{2}) \bigg) \nonumber\\
	&+& \frac{1}{H^4(z)} \bigg(2 c H (z) (z+1)^3-2 c (z+1)^4-H^2 (z)+3 H_0^2 (\Omega _{\text{$\Lambda $0}} - (z+1)^6 \Omega _{\text{$\sigma $0}}) \nonumber\\
	&+&3 H(z) (\xi_{0}+\xi_{1} H(z)+\xi_{2} (\dot H(z)+ H(z)^{2}) \bigg){}^2\nonumber\\
	&-&\frac{1}{2 H^2(z)}\bigg(2 c H (z) (z+1)^3-2 c (z+1)^4-H^2 (z)\nonumber\\
	&+&3 H_0^2 (\Omega _{\text{$\Lambda $0}}- (z+1)^6 \Omega _{\text{$\sigma $0}})+3 H(z) (\xi_{0}+\xi_{1} H(z)+\xi_{2} (\dot H(z)+ H(z)^{2})\bigg)\biggr\}\bigg]
\end{eqnarray}

\section{Hubble data from 57 observations, data set H (z)}
The observational data and statistical analytical analyses used to limit the model parameters of the generated universe are presented in this section (Figs. 1, 2). In this domain, we have applied 57 $H(z)$ observational data points ranges  $0\leq z\leq 2.36$ \cite{ref32a}, which were obtained by using the Monte Carlo Technique ( MCT)" \cite{ref32b}. We employ the $chi2$ statistic to identify the best-fitting values and bounds for a fitted model. 

\begin{equation}\label{33}
	\chi^{2}\left(p\right)=\sum_{i=1}^{57} {\frac{\left(H_{th}(i)-H_{ob}(i)\right)^2}{\sigma(i)^2}}
\end{equation}

The OHD measurements presented in Table 1 are consistent with the $chi^2$ expression in Eq. (\ref {33}). A data set of the $H(z)$ observed values for $z$ is presented in this table. Various cosmologists have obtained a possible error by using the differential age approach. Here, $p$ is the set of parameters for the model, where ($p=H_{0}, \Omega_{m0}, \Omega_{\Lambda0},\Omega_{\sigma0} $). The $\chi^{2}$ expression in  Eq. (\ref{33}) is true for all of the $H(z)$ OHD measurements in Table 1. This Table shows data about the observed values of the Hubble parameters $H(z)$ versus redshift $z$, along with a possible error. These values were found by different cosmologists using the various age approach.

\begin{table}[H]
	\caption{\small The behavior of Hubble parameter H(z) with redshift}
	\begin{center}
		  \begin{tabular}{|c|c|c|c|c|c|c|c||c|c|c|c|c|c|c|c|}
		  				  
			\hline
		``	\tiny	  $Z$ & \tiny $H (Obs)$ & \tiny $\sigma_{i}$ & \tiny Ref 
			& 	\tiny	  $Z$ & \tiny $H (Obs)$ & \tiny $\sigma_{i}$ & \tiny Ref 
			& 	\tiny	 $Z$ & \tiny $H (Obs)$ & \tiny $\sigma_{i}$ & \tiny Ref 
			&\tiny	  $Z$ & \tiny $H (Obs)$ & \tiny $\sigma_{i}$ & \tiny Ref\\ 
			\hline
			
			\tiny	  $0.07$ & \tiny 69  & \tiny  19.6 & \tiny \cite{ref72} 
			& 	\tiny	0.4783   & \tiny 80 & \tiny 90  & \tiny \cite{ref31} 
			& 	\tiny 0.24	  & \tiny 79.69 & \tiny 2.99  & \tiny \cite{ref78} 
			&\tiny	0.52   & \tiny 94.35  & \tiny 2.64  & \tiny \cite{ref78} \\ 
			\hline
			\tiny	  $0.90$ & \tiny 69  & \tiny 12  & \tiny \cite{ref73}
			& 	\tiny	0.480   & \tiny 97 & \tiny 62 & \tiny \cite{ref72}
			& 	\tiny	0.30  & \tiny 81.7 & \tiny 6.22 & \tiny \cite{ref79} 
			&\tiny	0.56   & \tiny 93.34  & \tiny 2.3 & \tiny \cite{ref52}\\ 
			\hline
			\tiny	  $0.120$ & \tiny 68.6  & \tiny26.2   & \tiny \cite{ref72}
			& 	\tiny	0.593   & \tiny 104 & \tiny 13 & \tiny \cite{ref84}
			& 	\tiny.31	  & \tiny 78.18 & \tiny 4.74 & \tiny \cite{ref52} 
			&\tiny.57	   & \tiny 87.6  & \tiny 7.8 & \tiny \cite{ref81}\\ 
			\hline
			\tiny	  $0.170$ & \tiny 83  & \tiny 8  & \tiny \cite{ref73} 
			& 	\tiny	0.6797   & \tiny 92 & \tiny 8 & \tiny \cite{ref31}
			& 	\tiny0.34	  & \tiny 83.8  & \tiny 3.66 & \tiny \cite{ref78} 
			&\tiny 0.57	   & \tiny 96.8 & \tiny  3.4 & \tiny \cite{ref82}\\ 
			\hline
			
			\tiny .1791	  & \tiny 75   & \tiny 4  & \tiny \cite{ref74}
			& 	\tiny.7812	   & \tiny 105 & \tiny 12 & \tiny \cite{ref74} 
			& 	\tiny 0.35	  & \tiny 82.7  & \tiny 9.1 & \tiny \cite{ref80}
			&\tiny 0.59   & \tiny 98.48 & \tiny 3.18  & \tiny \cite{ref52}\\ 
			\hline
			
			\tiny0.1993	  & \tiny  75  & \tiny  5 & \tiny \cite{ref74}
			& 	\tiny 0.8754	   & \tiny 124  & \tiny 17 & \tiny \cite{ref74} 
			& 	\tiny 0.36	  & \tiny 79.94 & \tiny 3.38 & \tiny \cite{ref52}
			&\tiny  0.60  & \tiny 87.9  & \tiny 6.1  & \tiny \cite{ref45}\\ 
			\hline
			
			\tiny0.200	 & \tiny  72.9   & \tiny 29.6  & \tiny \cite{ref75} 
			& 	\tiny 0.880	   & \tiny 90   & \tiny 40 & \tiny  \cite{ref72}
			& 	\tiny 0.38	  & \tiny 81.5 & \tiny 1.9 & \tiny \cite{ref47} 
			&\tiny 0.61    & \tiny 97.3 & \tiny 2.1  & \tiny \cite{ref47}\\ 
			\hline
			\tiny 0.270	   & \tiny 77   & \tiny  14 & \tiny \cite{ref73}
			& 	\tiny0.900	   & \tiny 117  & \tiny 23  & \tiny \cite{ref73}
			& 	\tiny 0.40	  & \tiny 82.04 & \tiny  2.03 & \tiny \cite{ref52}
			&\tiny 0.64   & \tiny 97.3  & \tiny 2.1  & \tiny \cite{ref52}\\ 
			\hline
			\tiny	 0.280 & \tiny  88.8   & \tiny 36.6   & \tiny \cite{ref31}
			& 	\tiny	1.037   & \tiny 154  & \tiny 20  & \tiny \cite{ref73} 
			& 	\tiny 0.43	  & \tiny 86.45 & \tiny 3.97 & \tiny \cite{ref76}
			&\tiny 0.73    & \tiny 97.3  & \tiny 7  & \tiny \cite{ref80}\\ 
			\hline
			\tiny	  0.3519 & \tiny  83  & \tiny 14  & \tiny \cite{ref74} 
			& 	\tiny1.300	   & \tiny 168 & \tiny  17 & \tiny \cite{ref73}
			& 	\tiny 0.44	  & \tiny 82.6  & \tiny 7.8  & \tiny \cite{ref52}
			&\tiny 2.30   & \tiny 224  & \tiny 8.6  & \tiny \cite{ref27}\\ 
			\hline
			
			\tiny	  0.3802 & \tiny 83   & \tiny 13.5  & \tiny \cite{ref31} 
			& 	\tiny 1.363	   & \tiny 160 & \tiny 33.6 & \tiny \cite{ref77} 
			& 	\tiny 0.44	  & \tiny 84.81  & \tiny 1.83 & \tiny \cite{ref52}
			&\tiny 2.33   & \tiny 224 & \tiny 8  & \tiny  \cite{ref83} \\
			\hline
			\tiny	  0.400 & \tiny  95  & \tiny 17  & \tiny \cite{ref73}
			& 	\tiny 1.430	   & \tiny 177 & \tiny 18  & \tiny \cite{ref73}  
			& 	\tiny0.48	  & \tiny 87.79  & \tiny 2.03  & \tiny \cite{ref52}
			&\tiny 2.34   & \tiny 222 & \tiny 8.5   & \tiny \cite{ref84}\\ 
			\hline
			\tiny	0.4004 & \tiny 77   & \tiny 10.2  & \tiny \cite{ref31}
			& 	\tiny 1.530	   & \tiny 140 & \tiny 14 & \tiny \cite{ref73}
			& 	\tiny 0.51	  & \tiny 90.4 & \tiny 1.9 & \tiny \cite{ref47}
			&\tiny 2.36  & \tiny 226 & \tiny  9.3 & \tiny \cite{ref85}\\ 
			\hline
			\tiny	  0.4247 & \tiny 87.1   & \tiny  11.2 & \tiny \cite{ref31}
			& 	\tiny1.750	   & \tiny 202 & \tiny 40 & \tiny \cite{ref73} 
			& 	\tiny	  & \tiny  & \tiny  & \tiny 
			&\tiny    & \tiny  & \tiny   & \tiny \\ 
			\hline
			\tiny	0.4497 & \tiny 92.8   & \tiny 12.9  & \tiny \cite{ref31}
			& 	\tiny 1.965	   & \tiny 186.5  & \tiny 50.4 & \tiny \cite{ref77}
			& 	\tiny	  & \tiny  & \tiny  & \tiny 
			&\tiny    & \tiny  & \tiny   & \tiny \\ 
			\hline
			\tiny	 0.470 & \tiny  89  & \tiny 34  & \tiny \cite{ref76}''
			& 	\tiny	   & \tiny  & \tiny  & \tiny  
			& 	\tiny	  & \tiny  & \tiny  & \tiny  
			&\tiny    & \tiny  & \tiny   & \tiny  \\ 
			\hline

		\end{tabular}
	\end{center}
\end{table}
\begin{figure}[H]
	{\centering
		\includegraphics[width=8cm,height=8cm,angle=0]{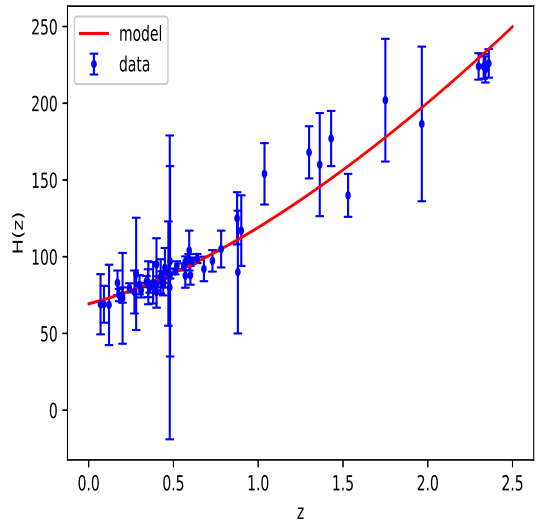}
	\caption{The figure shows the comparison of theoretical model with error bar plots of the OHD points}}
\end{figure} 

\begin{figure}[H]
	{\centering
	\includegraphics[width=13cm,height=13cm,angle=0]{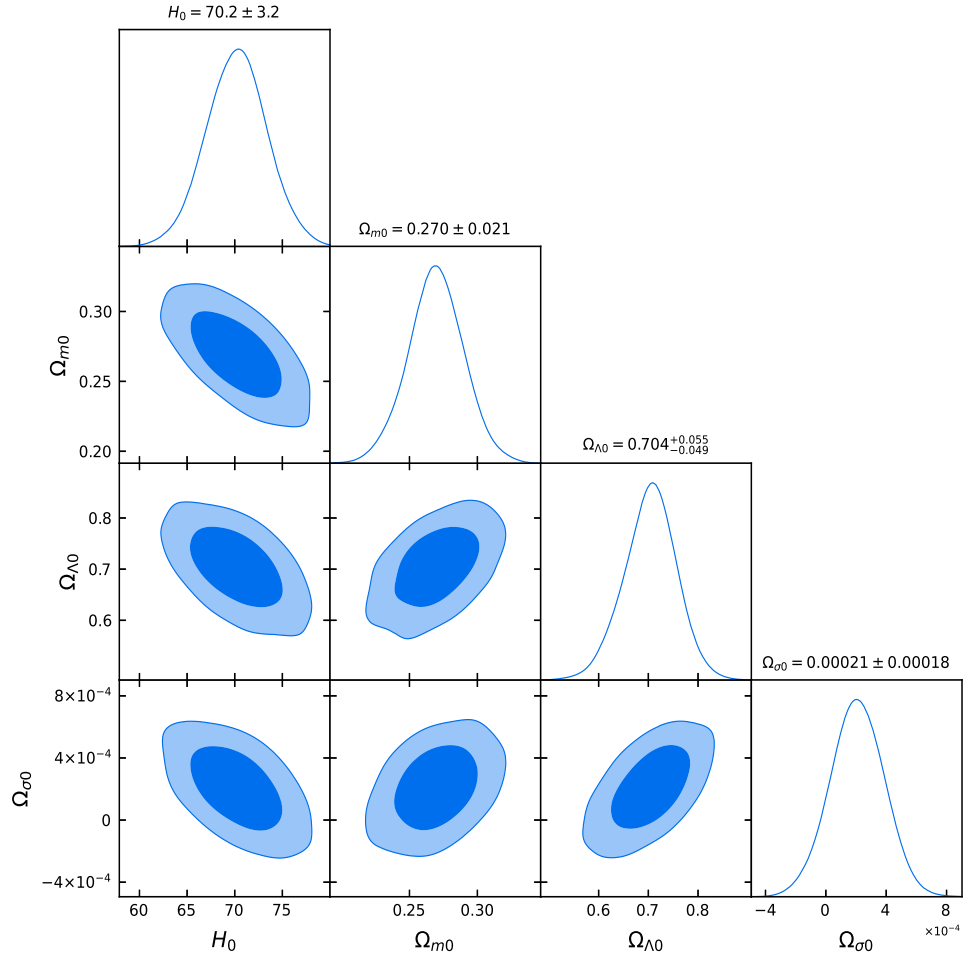}
		\caption{Plots for 1-D marginalized distribution, and 2-D contours with 68$\%$ CL and 95$\%$ CL for the model parameters. Marginalized probability distributions of the individual parameters are also displayed.}}
\end{figure} 

\begin{figure}[H]
	{\centering
		(a)\includegraphics[width=7cm,height=5cm,angle=0]{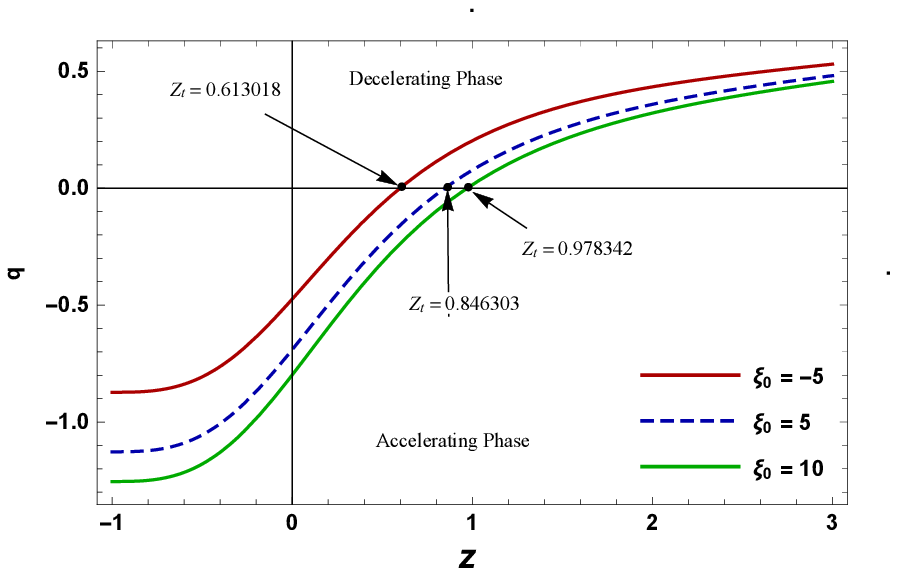}
		(b)\includegraphics[width=7cm,height=5cm,angle=0]{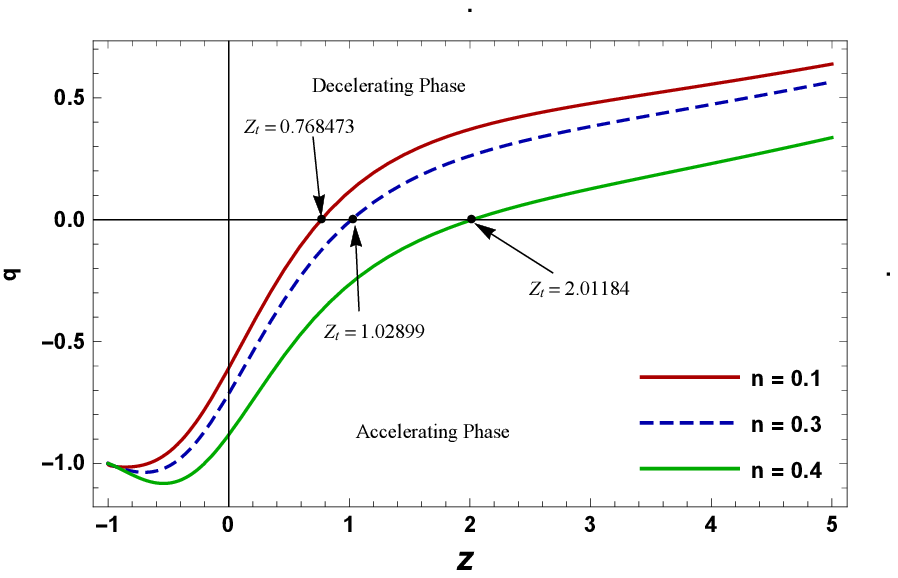}
		(c)\includegraphics[width=7cm,height=5cm,angle=0]{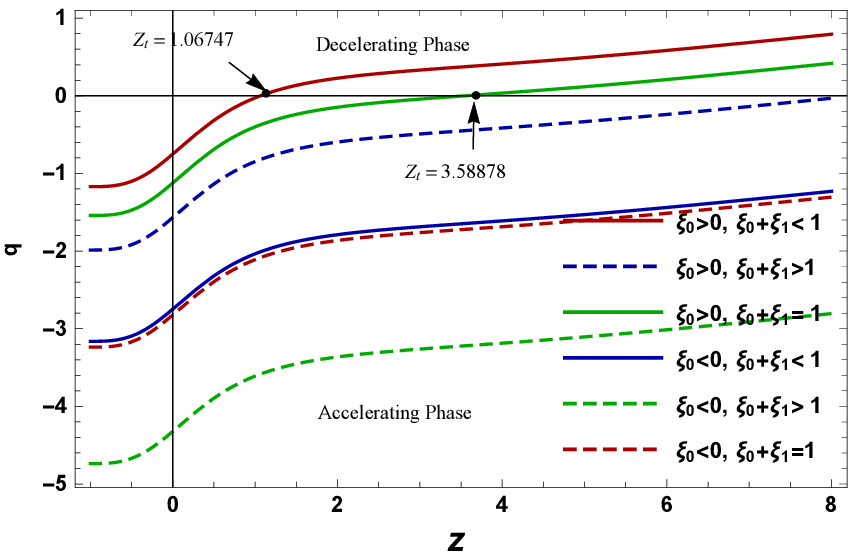}		(d)\includegraphics[width=7cm,height=5cm,angle=0]{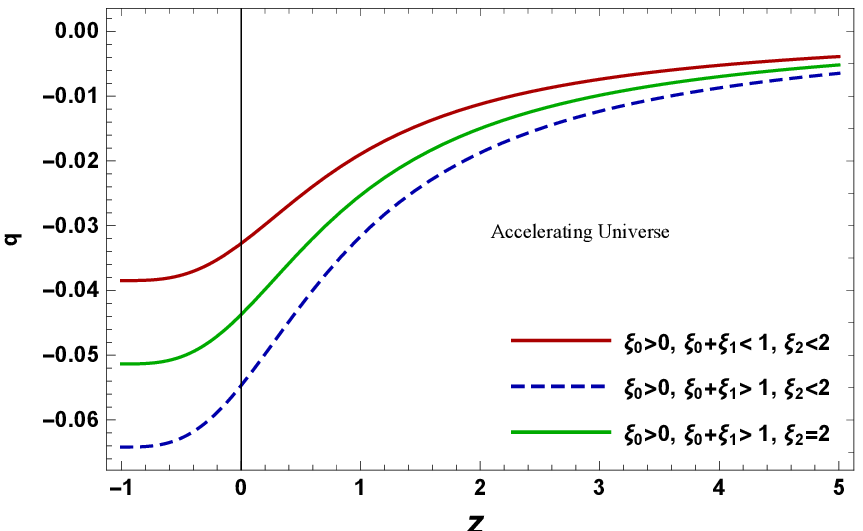}
		\caption{ The plots show the behavior of deceleration parameter $q$ versus $z$ for different scenarios. Here $H_{0} = 70.2$,  $\Omega_{m0} = 0.270$, $\Omega_{\Lambda_{0}} = 0.704$, $\Omega_{\sigma_{0}}  0.00021$, and c=0.001.}}
\end{figure} 


\begin{figure}[H]
{\centering}
	(a)\includegraphics[width=7cm,height=5cm,angle=0]{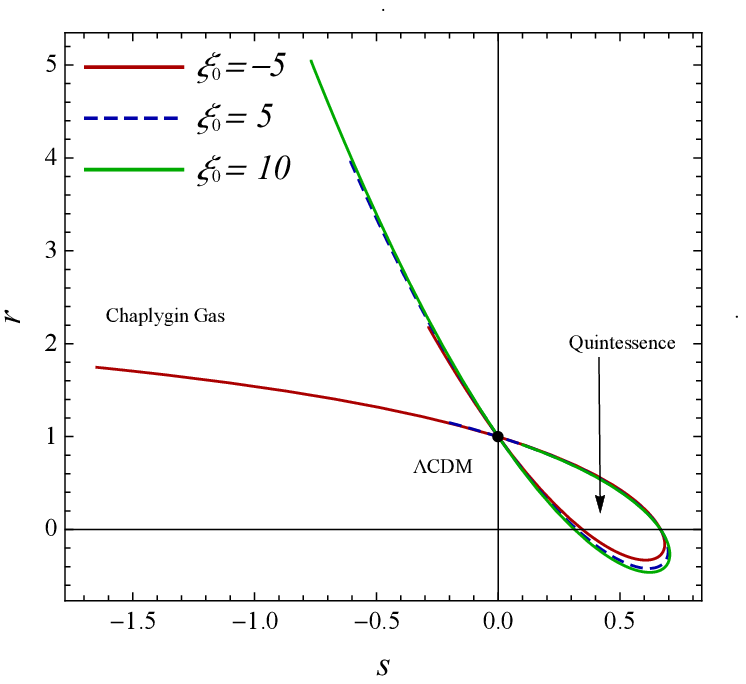}
	(b)\includegraphics[width=7cm,height=5cm,angle=0]{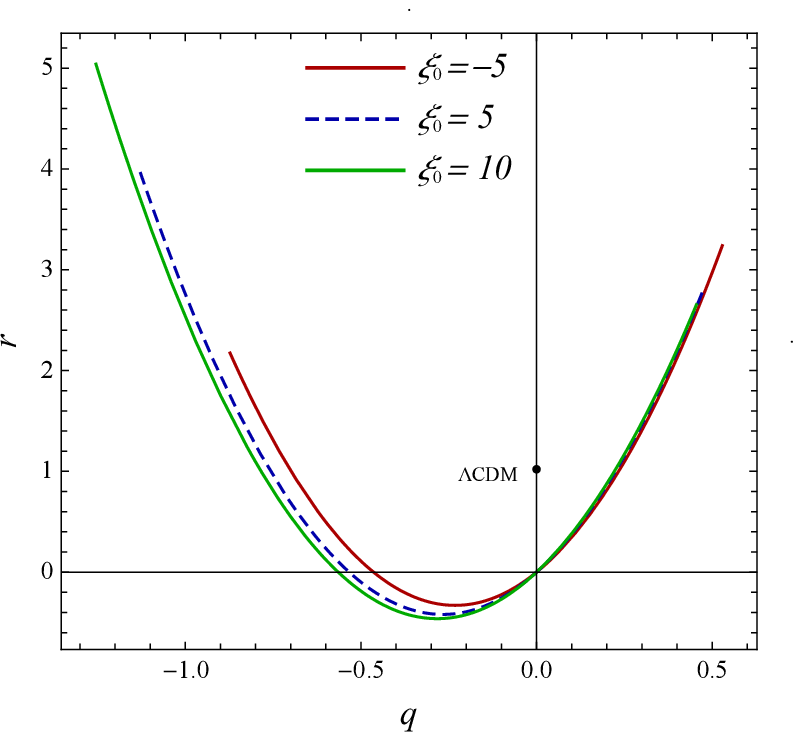}
	(c)\includegraphics[width=7cm,height=5cm,angle=0]{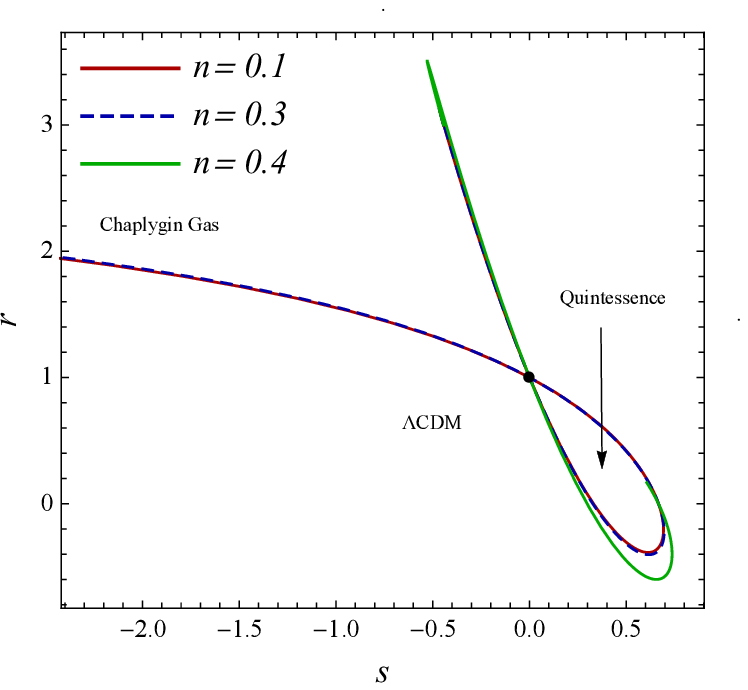}
	(d)\includegraphics[width=7cm,height=5cm,angle=0]{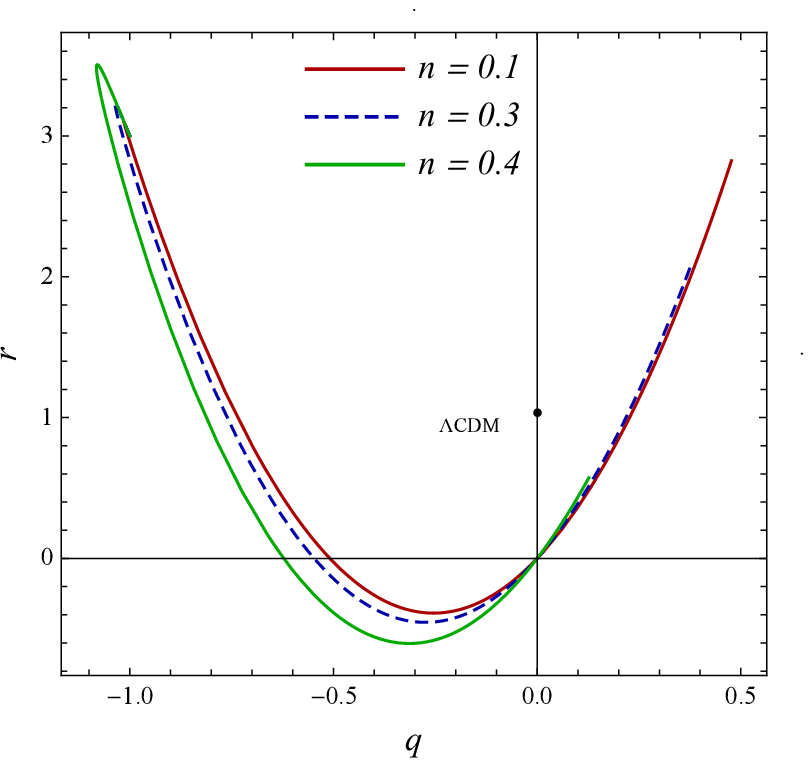}
    (e)\includegraphics[width=7cm,height=5cm,angle=0]{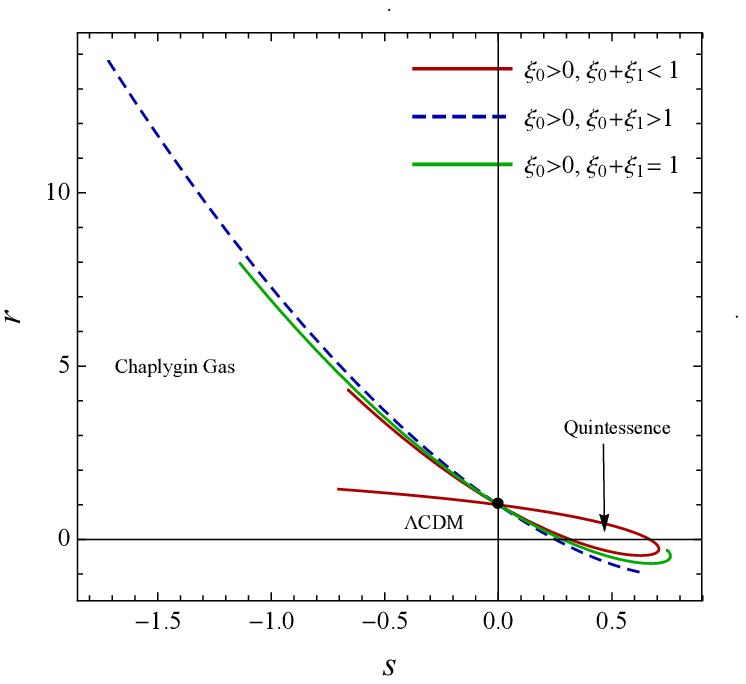}
    (f)\includegraphics[width=7cm,height=5cm,angle=0]{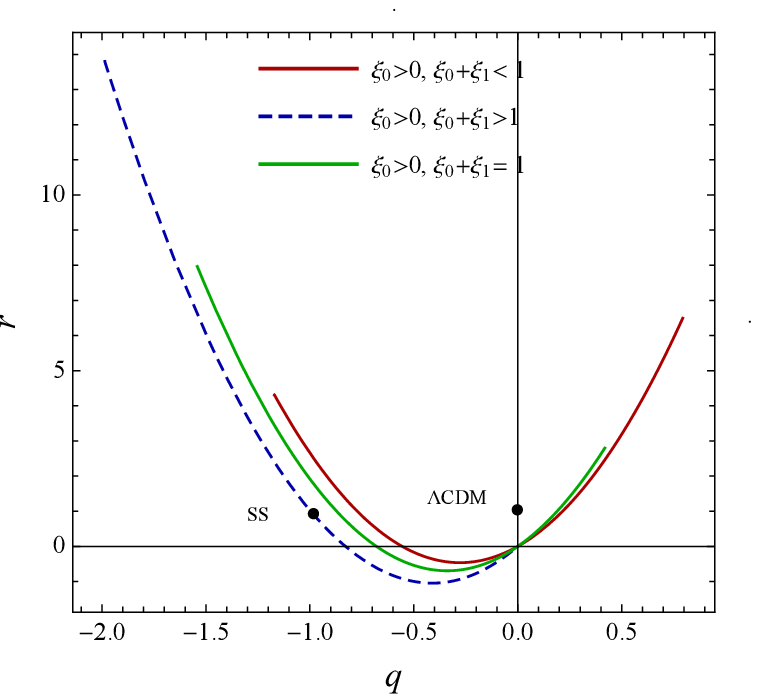}
    (g)\includegraphics[width=7cm,height=5cm,angle=0]{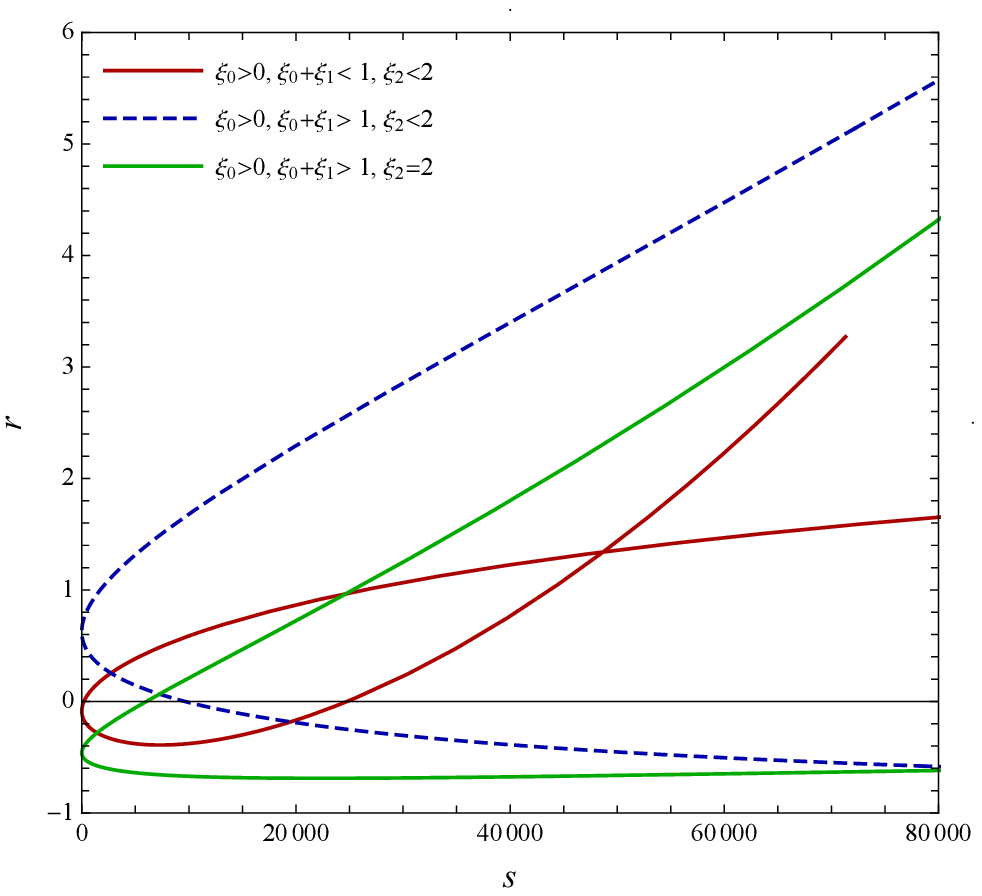}
    (h)\includegraphics[width=7cm,height=5cm,angle=0]{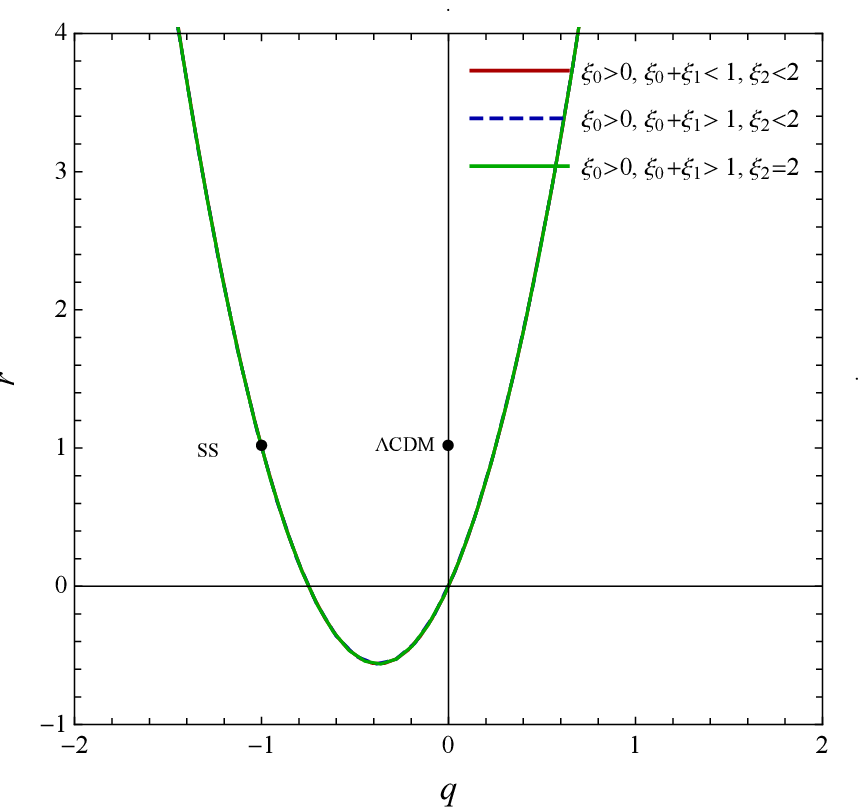}
	\caption{
	 The variations of $r$ versus $s$ are shown in ( figures 4a, 4c, 4e $\&$ 4g).
	 The black dot denotes the statefinder pair $(r, s) = (1, 0)$, or the position of a flat $\Lambda$CDM model in the $(r-s)$ plane. The location of statefinder parameters in the best-fitting model based on 57 H(z) data is shown by the star symbol on the $(r-s)$ curve.
	The variations of $r$ versus $q$ are shown in ( figures 4b, 4d, 4f $\&$ 4h). The black dot represents the location of SS point $(r, q) = (1, -1)$ in the $r–q$ plane,
 In both panels, we have used four bulk viscosity parameterizations, here
 $H_{0} = 70.2$, $\Omega_{m0} = 0.270$, $\Omega_{\Lambda_{0}} = 0.704$,
 $\Omega_{\sigma_{0}} = 0.00021$, and $c=0.001.$}
	\end{figure} 
\section{Result and Dissussion}

We use the 57 OHD points to constrain the free parameters of the model under consideration and perform the MCMC technique using the EMCEE Python module. We determined the best-fit parameters in our derived model by minimising the $\chi^{2}$ for the aforementioned datasets. The relationship between $q$ and $z$ for $\xi = -5, 5$ and $10$ is seen in figure 3a. The model demonstrates that the display transitioned smoothly from early $q>0$ to late $q < 0 $. According to the OHD results, the value of $q$ that is accelerating at the moment is in the range $-1\leq q<0$, which implies a decelerating universe, whereas its ``negative" value results in an accelerated universe. The results of the 57 data points in the observed high redshift OHD show that our model transits from an early decelerating to a late accelerating cosmos. For different values of $\xi= -5, 5, or 10$, we discovered that the early Universe was expanding with positive to negative values of $q$ and that there was a signature flipping at $z_{t} =0.80755.$, $z_{t} =0.94572$ and $z_{t} =1.06415$.  We can also observe that $q$ will become -1 in the future when $z$ reaches that value. The DP is a function of redshift. The Universe evolves to the infinite future $(z=-1)$ from an inflationary stage at the infinite past $(z=0)$. The K-matter solution (q = 0) appears to be the centre of the universe's dynamical plane (DP). We can see from Table-2 above that the Universe normally begins with a slowing down phase before transitioning to a speeding up phase for various values of $\xi$.

\begin{table}[H]
	\caption{\small The behaviour of the evoluation of the universe with transitioning point. }
	\begin{center}
		
		\begin{tabular}{| c| c|c|c|}	
			\hline
			range of $\xi$ &     Nature of $q$  &$z_{t}$  &Evoluation of the Universe   \\
			[0.5ex]
			\hline
			
		``	$-10$ & +ve to -ve  &0 .59043  &  decelerating-  Acclerating \\		
			$-5$ &  +ve to -ve  &  0.70886  &  decelerating-  Acclerating   \\
			$0$ & +ve to -ve     & 0.82792    & decelerating-  Acclerating     \\
			$5$ & +ve to -ve   & 0.94572  &  decelerating-  Acclerating  \\
			$10$ &+ve to -ve ''   &1.06415   &  decelerating-  Acclerating   \\
			\hline		
		\end{tabular}\\
	\end{center}
\end{table}
Figure 3b shows the nature of $q$ with respect to redshift $z$  for $0\leq n\leq 1/2$. The above Table-3 shows that in various values of $n$, the Universe begins smoothly with the deceleration phase to the accelerating phase. We observe that in the future, when z$\to$ -1, q$\to$ -1.


\begin{table}[H]
	\caption{\small The behaviour of the evoluation of the universe with transitioning point.}
	\begin{center}
		\begin{tabular}{| c| c|c|c|}	\hline
			
			range of n &     Nature of $q$  &$z_{t}$  &Evoluation of the Universe   \\
			[0.5ex]
			\hline
			
		``	$0.1$ & +ve to -ve  & 0.85553  &  decelerating -Acclerating   \\		
			$0.2$ & +ve to -ve   & 0.93313   &  decelerating -Acclerating \\
			$0.3$ &  +ve to -ve & 1.0933    &  decelerating -Acclerating   \\
			$0.4$ &  +ve to -ve  &1.6051   &   decelerating -Acclerating \\
			$0.5$ &  +ve to -ve''  & 1.9079  &  decelerating -Acclerating    \\
			\hline
			
		\end{tabular}\\
	\end{center}
\end{table}

Figure 3c shows  the nature of $q$ against redshift $z$  for the limiting conditions of the viscous parameters  with $\xi_{0} +\xi_{1}>1$,$\xi_{0} +\xi_{1}<1$ and $\xi_{0} +\xi_{1}=1$. The current deceleration parameter $q_{0} >0$  if $\xi_{0} + \xi_{1}> 1$, for $\xi_{0}>0$ and $\xi_{1}>0$ indicates that the universe is currently in the decelerating epoch and will transition to the accelerating phase at some point in future if  $\xi_{0}<0$ (see Table 4). We have different types of expansions of the Universe:
\begin{itemize}
\item For $q > 0$, the expansion of the universe would be deceleration.
\item For $q = 0$, it would expand at a constant rate.
\item If $-1<q>0$, the universe would be accelerating with power-law expansion.
\item If $q=-1$, the universe has exponential expansion (also known as de Sitter-expansion)
\item If $q<-1$, the universe would be super-exponential expansion.
\end{itemize}
If the  limiting conditions of the viscous parameters  with $\xi_{0}<0$ , our model shows only the deceleration phase, no trasition point obtained. In particular, the kinematic method of cosmic data analysis offers clear evidence for the universe's current acceleration stage, which is independent of the universe's matter-energy composition and the viability of general relativity.


\begin{table}[H]
	\caption{The behaviour of the evoluation of the universe with transitioning point.}
	\begin{center}
		
		\begin{tabular}{| c| c |c|  c| c|  c|}	\hline
			range of $\xi$ &  $\xi_{0}$ & $\xi_{1}$ &  Nature of $q$ & $z_{t}$  & Evolution of Universe \\
			[0.5ex]
			\hline
			
		``	$\xi_{0}>0$, $\xi_{0}$+$\xi_{1}<1.0$ & 0.8 & 0.1 & +ve to -ve  &1.4536   & decelerating -Acclerating \\
			
			$\xi_{0}>0$, $\xi_{0}$+$\xi_{1}>1.0$ & 0.45& 0.65& +ve to -ve  & 3,5882 &  decelerating -Acclerating \\
			
			$\xi_{0}>0$, $\xi_{0}$+$\xi_{1}=1.0$ & 0.65  &0.35 & +ve to -ve  &7.0663 &  decelerating -Acclerating \\
			$\xi_{0}<0$, $\xi_{0}$+$\xi_{1}<1.0$  &-0.5  & 1.45&   -ve   &-  &  Acclerating   \\
			$\xi_{0}<0$, $\xi_{0}$+$\xi_{1}>1.0$  & -0.5   &  2.5&  -ve  & -  &  Acclerating \\
			$\xi_{0}<0$, $\xi_{0}+\xi_{1}=1.0$''&  -0.5   & 1.5 &  -ve & - &  Acclerating   \\
			\hline
		\end{tabular}
	\end{center}
\end{table}

Similarly, Figure 3d shows the nature of $q$ against redshift $z$  for the limiting conditions of the viscous parameters $\xi_{0} +\xi_{1}>1$, $\xi_{0} +\xi_{1}<1$ and $\xi_{0} +\xi_{1}=1$. We obtained the current deceleration parameter only in the accelerating phase for all $\xi_{0}$, $\xi_{1}$, and $\xi_{2}$. The negative sign indicates inflation (see Table 5).

 
 \begin{table}[H]
 	\caption{The behaviour of the evoluation of the universe with transitioning point.}
 	\begin{center}
 		\begin{tabular}{| c| c |c|  c| c|  c|c|}	\hline
 			range of $\xi$ &  $\xi_{0}$ & $\xi_{1}$ &  $\xi_{2}$ &  Nature of $q$  &$z_{t}$  & Evolution of the universe \\
 			[0.5ex]
 			\hline
 			
 			``$\xi_{0}>0$, $\xi_{0}$+$\xi_{1}<1.0, \xi_{2}<2.0$ & 0.9 & .01 & 1& -ve  &- & Acclerating\\
 			
 			$\xi_{0}>0$,$\xi_{0}$+$\xi_{1}>1.0, \xi_{2}<2.0$  & 0.45 & 0.65 & 1&-ve &-& Acclerating  \\
 			
 			$\xi_{0}>0$,$\xi_{0}$+$\xi_{1}=1.0, \xi_{2}<2.0$  &0.65  &.35 & 1& -ve &- & Acclerating   \\
 			$\xi_{0}>0$, $\xi_{0}$+$\xi_{1}<1.0, \xi_{2}>2.0$  &  -0.5  &1.45 & 2.1 & -ve &- & Acclerating   \\
 			$\xi_{0}<0$,$\xi_{0}$+$\xi_{1}>1.0, \xi_{2}>2.0$  & -0.5   &  2.5&3 &-ve &- & Acclerating \\
 			$\xi_{0}<0$,$\xi_{0}+\xi_{1}>1.0, \xi_{2}>2.0$ '' &  -0.5   & 1.5 & 2.17 & -ve &- & Acclerating  \\
 			\hline
 			
 		\end{tabular}\\
 	\end{center}
 \end{table}


The $r-s$ trajectories in Fig. 4a are separated into two sections by a vertical line. According to references\cite{ref69,ref70}, the region $r> 1$, $s < 0$ in the $r-s$ plane exhibits behavior resembling a Chaplygin gas (CG) model, while the region  $r <1$, $s>0$ exhibits behavior resembling the quintessence model (Q-model). The trajectories in all regions coincide for all values of viscous parameter $\xi$, whether positive or negative. The current values of ${r, s}$ (to meet) the $\Lambda$ CDM point for various  values of $\xi_{0}$". Fig 4b shows the behavior of the model in the $(r-q)$ plane. The trajectories are plotted for $\xi=-5,5,10$, and the dot represents the $\Lambda$ CDM point (see Table-6, case I).


\begin{table}[H]
	\caption{A numerical description of cosmological model parameters}
	\begin{center}
		\begin{tabular}{| c| c|c|c|}	
			\hline	range of $\xi$ & $q$  & r  & s \\
			[0.5ex]
			\hline
			$-10$ 	& -0.203426 & -0.281664& 0.607344 \\
			$-5$ &-0.302147 & -0.299041& 0.539818 \\
			$0$ &-0.400869& -0.23845& 0.458243 \\
			$5$  &-0.49959& -0.099892 & 0.366781 \\
			$10$ &-0.598311 & 0.116633& 0.268098 \\
			\hline
		\end{tabular}
	\end{center}
\end{table}
In the case of II, our II parameterization is based on the density parameter. In this case, Fig. 4c  also shows the behavior similar to a CG region, whereas the region $r <1$, $s>0$ depicts the Q-model for $0\leq n<1/2$. The trajectory $r-s$ shows the Q-model early and approaches $\Lambda$CDM at late times. It also shows the CG at early times and approaches $\Lambda$ CDM at late times. Fig. 4d depicts that our model approaches $\Lambda$CDM model. All the trajectories coincide for all different values of $\xi$ (see Table-7, case II).


\begin{table}[H]
	\caption{A numerical description of cosmological model parameters}
	\begin{center}
		\begin{tabular}{| c| c|c|c|}	\hline
			range of n & $q$  & r  & s \\
			[0.5ex]
			\hline
			
			0.1 &-0.639804 & 0.230912 & 0.224918 \\
			0.2 	&-0.979166 & 1.68256 & -0.153816  \\
			0.3 	&-1.80053 & 9.00947 & -1.16053 \\
			0.4 	&-3.78847 & 49.0818 & -3.73729 \\
			0.5  &-8.5999 & 276.928 & -10.1074\\
			\hline
		\end{tabular}\\
	\end{center}
\end{table}

 The case-III shows  the nature of $r-s$ and $r-q$  for the limiting conditions of the viscous parameters  with $\xi_{0} +\xi_{1}>1$,$\xi_{0} +\xi_{1}<1$ and $\xi_{0} +\xi_{1}=1$. The current deceleration parameter $q_{0} >0$  if $\xi_{0} + \xi_{1}> 1$, for $\xi_{0}>0$ and $\xi_{1}>0$  shown in Fig 4e. $\&$ Fig. $4f$ The trajectory shows the Q-model behavior at early times for $\xi_{0}>,\xi_{1}>0$ and approaches $\Lambda$CDM at late times. In both figures the ``dot represents the fixed points {(r, q)} = {(1, 0)} and {(r, q)} = {(1,-1)} of LCDM and Steady State (SS) models", respectively. It can be observed that there is a sign change of $q$ from positive to negative in a Q-region at small values of $\xi_{0}$ and $\xi_{1}$. The q-factor has always negative values starting from $q<-1$ and tends to $q=-1$ at late times for large values of $\xi_{0}$ and $\xi_{1}$ (see Table-8, case-III).

 
\begin{table}[H]
	\caption{ A numerical description of cosmological model parameters}
	\begin{center}	
		\begin{tabular}{| c| c |c|  c| c|  c|}	\hline
			range of $\xi$ &  $\xi_{0}$ & $\xi_{1}$   &$q$  &$r$  & $s$ \\
			[0.5ex]
			\hline
			
		``	$\xi_{0}>0$, $\xi_{0}+\xi_{0}<1.0$ & 0.8 & 0.1 &-0.56666 & 0.03873 & 0.30039\\
			
			$\xi_{0}>0$, $\xi_{0}$+$\xi_{1}>1.0$  &0.45&0.65 &-1.38475 & 4.62605 & -0.64129  \\ 
			
			$\xi_{0}>0$, $\xi_{0}$+$\xi_{1}=1.0$  &   0.65  &0.35 & -0.93870 & 1.46109 & -0.10683  \\ 
			$\xi_{0}<0$, $\xi_{0}$+$\xi_{1}<1.0$ &  -0.5  & 1.45& -2.5659 & 20.6965 & -2.1413\\
			$\xi_{0}<0$, $\xi_{0}$+$\xi_{1}>1.0$ & -0.6   &  2.5& -4.13902 & 59.42716 & -4.19823 \\
			$\xi_{0}<0$, $\xi_{0}+\xi_{1}=1.0$ '' &  -0.7   & 1.7 & -2.93704 & 28.0484 & -2.62322  \\
			
			\hline		
		\end{tabular}
	\end{center}
\end{table}

Similarly, Fig. $4g$ depicts the evolution of our model in the $r-s$ plane. The $r-s$  and $r-q$ planes the trajectory shows the Q-model behavior at early times for  and approaches $\Lambda$CDM at late times for the  limiting conditions of the viscous parameters   $\xi_{0} +\xi_{1}>1$,$\xi_{0} +\xi_{1}<1$ and $\xi_{0} +\xi_{1}=1$, for $\xi_{0}>0,\xi_{1}>0,\xi_{2}>0$. In Fig. 4h the region $r>1$, $q<-1$ in the $r-q$ plane shows a behavior similar to the phantom model (see Table-9, case-IV).


\begin{table}[H]
	\begin{center}
		\caption{  A numerical description of cosmological model parameters}
		\begin{tabular}{| c| c |c|  c| c|  c|c|}	\hline
			range of $\xi$ &  $\xi_{0}$ & $\xi_{1}$ &  $\xi_{2}$ &  $q$  & $r$ & $s$ \\
			[0.5ex]
			\hline
			
		``	$\xi_{0}>0$, $\xi_{0}$+$\xi_{1}<1.0, \xi_{2}<2.0$ & 0.9 & .01 & 1&-0.020112& 8753.38 & -61.348\\
			
			$\xi_{0}>0$,$\xi_{0}$+$\xi_{1}>1.0, \xi_{2}<2.0$  & 0.45 & 0.65 & 1&-0.020113 & 8756.63& -61.360 \\
			
			$\xi_{0}>0$,$\xi_{0}$+$\xi_{1}=1.0, \xi_{2}<2.0$  &0.65  &.35 & 1& -0.020036 & 8590.46& -60.7656\\
			$\xi_{0}>0$, $\xi_{0}$+$\xi_{1}<1.0, \xi_{2}>2.0$  &  -0.5  &1.45 & 2.1 &-0.04218 & 38586.56 & -129.9275  \\
			$\xi_{0}<0$,$\xi_{0}$+$\xi_{1}>1.0, \xi_{2}>2.0$  & -0.5   &  2.5&3 & -0.06033& 79415.51& -186.841 \\
			$\xi_{0}<0$,$\xi_{0}+\xi_{1}>1.0, \xi_{2}>2.0$''  &  -0.5   & 1.5 & 2.17 & -0.04358 & 41200.78 & -134.290  \\
			\hline
			
			\hline	
		\end{tabular}\\
	\end{center}
\end{table}


\section{Conclusion}
As we all know, bulk viscosity is a very important part of cosmology. It helps explain why the universe expands quickly during the inflationary phase. 
The impact of viscosity on the evolution of cosmological models and the role of viscosity in avoiding the first big bang singularity have been examined by a number of authors \cite{ref13,ref14,ref15}.
In this study, we examine the Bianchi type-I model with bulk viscosity and the role of bulk viscosity in explaining the early and late universe. The model incorporates the ideal fluid with the four forms of bulk viscosity parameterization. We have performed a detailed study of the Bianchi-I viscous model. We have  estimated values of cosmological parameters from the OHD data sets. We have studied four cases of parameterization.\\

We have provided a summary of the key findings:
\begin{itemize}
\item 
Fig. 1 shows the 2-dimensional joint contours (57 OHD), at 68$\% $and 95$\% $ confidence regions,  bounded with latest. We have find the  constraint  on $H_{0}$, $\Omega_{\text{m0}}$ $\Omega_{\Lambda_{0}},$$\Omega_{\sigma_{0}}$ by (57 OHD ) data sets (Table-I). The best fit values $H_{0}$ are closer to other investigation.

\item 
 The error bar plots for OHD data sets are shown in Fig. 2. The dots in both graphs reflect the observed values after corrections, and the red lines compare the current model to the $\Lambda$CDM model, which is depicted by the black dashed lines.

\item 

As shown in the graph, the deceleration parameter $q$ oscillates with $z$ from positive to negative values. The generated model describes the transition from the early deceleration phase to the current expanding phase of the universe. Various viscosity parameterizations for cases I, II, and III are displayed in Figs. $3a, $3b, $3c, and $3d.
We also observed that q$\to$ -1 in the future when z$\to$ -1. In case of IV, we obtained  the current deceleration parameter only in accelerating phase for all $\xi_{0}$, $\xi_{1}$ and  $\xi_{2}$. We can say that the plot of $q$ versus $z$ exhibits cosmic deceleration for high redshift $z$, acceleration for low redshift. We observe that variation of $q$ with bulk viscous coefficient  $\xi$ and the model paramters $H_{0} =70.2\pm3.2$, $\Omega_{\text{m0}}=.270\pm0.021$, $\Omega_{\Lambda_{0}}=0.704^{+0.055}_{-0.049},$ $\Omega_{\sigma_{0}}=0.00021\pm0.00018$  have shown  in Table $2$, $3$ , $4$ and $5$. The dynamics of the deceleration parameter exhibits a signature transitioning from an early phase of decelerating to the phase in which it is now accelerating. Different studies show transition redshifts as  $z_{t} = {0.69}^{+0.09}_{-0.06}, {0.65}^{+0.10}_{-0.07}$ and ${0.61}^{+0.12}_{-0.08}$ within  $(1 \sigma)$ for the joined $(SNIa + CC + H_{0})$ data, $z_{t}={0.646}^{+0.020}_{-0.158}$, $z_{t}={0.659}^{+0.371}_{-0.124}$, $z_{t}={0.860}^{+0.013}_{-0.146}$, $z_{t}={1.183}^{+0.002}_{-0.032}$ for SN, OHD and BAO data and $z_{t}={0.8596}^{+0.2886}_{-0.2722}$, $z_{t}={0.6320}^{+0.1605}_{-0.1403}$, and $z_{t}={0.7679}^{+0.1831}_{-0.1829}$ for the models using SNIa, OHD and combined data \cite{ref86,ref87,ref88,ref89,ref90}. Thus, results in the present manuscript show a agreement with the recent analysis.

\item 
We also observed the $(r~s)$ and $(r~q)$ plane parameters to diagnose the DE model geometrically, as shown in Figs. 4a,4c,4e $\&$ 4f . Our derived model initially shows a Chaplygin gas (CG) type DE model, which later evolves into a quintessence DE model at a few points. The model, later on, reverts again to CG. Interestingly, the model deviates significantly from the point $(r~s)= (1,0)$, which coincides with $\Lambda CDM$ see Figs. 4b, 4d, 4e $\&$ 4g for four viscosity parameterization.The detail study of trajectories are shown in Tables 6, 7, 8, 9 for the constrained value of the model parameters $H_{0} =70.2\pm3.2$, $\Omega_{\text{m0}}=.270\pm0.021$, $\Omega_{\Lambda_{0}}=0.704^{+0.055}_{-0.049},$ $\Omega_{\sigma_{0}}=0.00021\pm0.00018$. 
\end{itemize}
We have now come to the conclusion that bulk viscosity can be used to explain DE phenomena.
\section*{Acknowledgments}
The author (A. Pradhan) is grateful for the assistance and facilities provided by the University of Zululand, South Africa during a visit where a part of this article was completed.



\end{document}